\documentclass[groupedaddress,aps,pra,superscriptaddress,showpacs,twocolumn,prl]{revtex4}%
\usepackage{epsfig,dsfont,amssymb,amsmath,amsthm,amsfonts,amsbsy,mathrsfs}
\usepackage{graphicx, color}
\usepackage{epstopdf}
\usepackage{mathdots}
\usepackage{xcolor}
\usepackage{float}
\begin{document}

\title{Quantum Coherence Bounds the Distributed Discords}

\author{Zhi-Xiang Jin}
\thanks{Corresponding author: zhijin@mis.mpg.de}
\affiliation{School of Physics, University of Chinese Academy of Sciences, Yuquan Road 19A, Beijing 100049, China}
\affiliation{Max-Planck-Institute for Mathematics in the Sciences, Leipzig 04103, Germany}
\author{Xianqing Li-Jost}
\thanks{Corresponding author: xianqing.li-jost@mis.mpg.de}
\affiliation{Max-Planck-Institute for Mathematics in the Sciences, Leipzig 04103, Germany}
\author{Shao-Ming Fei}
\thanks{Corresponding author: feishm@cnu.edu.cn}
\affiliation{School of Mathematical Sciences,  Capital Normal University,  Beijing 100048,  China}
\affiliation{Max-Planck-Institute for Mathematics in the Sciences, Leipzig 04103, Germany}
\author{Cong-Feng Qiao}
\thanks{Corresponding author: qiaocf@ucas.ac.cn}
\affiliation{School of Physics, University of Chinese Academy of Sciences, Yuquan Road 19A, Beijing 100049, China}
\affiliation{CAS Center for Excellence in Particle Physics, Beijing 100049, China\\ \vspace{7pt}}

\begin{abstract}
Establishing quantum correlations between two remote parties by sending an information carrier is an essential step of many protocols in quantum information processing. We obtain trade-off relations between discords and coherence within a bipartite system. Then we study the distribution of coherence in a bipartite quantum state by using the relations of relative entropy and mutual information. We show that the increase of the relative entropy of discord between two remote parties is bounded by the nonclassical correlations quantified by the relative entropy of coherence between the carrier and two remote parties, providing an optimal protocol for discord distribution and showing that quantum correlations are the essential resource for such tasks.
\end{abstract}
\maketitle

\section{Introduction}
Quantum coherence and quantum correlations like quantum discord are valuable resources in quantum information processing \cite{horm,modi,hml,lipp}. Stemming from the superposition rule of quantum mechanics, quantum coherence captures the feature of quantumness in a single system and plays an important role in a variety of applications ranging from thermodynamics \cite{mkd,mdt} to metrology \cite{vsl}.
In Ref. \cite{hml}, the authors provide a full review about the resource theory of quantum coherence, including its application in many-body systems, and the discordlike quantum correlations.
Recently, the resource theory of coherence has attracted much attention, with efforts to the quantification and manipulation of coherence \cite{winter,jaberg,gour,tmm,agm}. 
Intrestingly, in Ref. \cite{gl}, the author present the quantum correlation measure equivalence in dynamic causal structures of quantum gravity, which maybe helpful for investigating the quantification of coherence.
Coherence in multipartite systems has been also studied \cite{yys,crad,asmm}, together with its relations to quantum entanglement and quantum nonlocality \cite{mnb1,mnb2,ecm,zhuhj,xiya}.
The distribution of coherence in bipartite and multipartite systems has been investigated in Ref. \cite{mt} and \cite{crad}, respectively. In \cite{crad} the trade-off relation between the intrinsic coherence and the local coherence in multipartite systems has been demonstrated. In \cite{tkc,ahd} the authors proved that the increase of relative entropy of entanglement between two remote parties is bounded by the amount of nonclassical correlations. A rigorous characterization of the distribution of coherence in multipartite systems is imperative and of paramount importance.

The quantum discord quantifies the quantum correlation in a bipartite systems and plays a central role in quantum tasks due to its potential applications in such as quantum critical phenomena \cite{mss,wte,jma,lsw} and quantum evolution under decoherence \cite{sha,jma1,aaf,lma}.
We address the following fundamental questions: How much can the discord between sender and receiver laboratories increase under the exchange of a carrier? Is there a quantitative relation between such increase and the nonclassical correlations between the carrier and the parties?

In this article, we present a general bound on the discord gain between distant laboratories under local quantum-incoherent operations and quantum communication, which is given by the quantum coherence between them and the carrier.  We first give some trade-off relations between various types of discord and coherence within a bipartite system.
Then, we discuss the distribution of coherence in a bipartite quantum state into discord between subsystems and coherent in each individual subsystem, by using the relations of relative entropy and mutual information. Finally, discord distribution in multipartite state is studied, and the discord gain between distant laboratories is bounded by the amount of quantum coherence between them and the carrier.

\section{Linking quantum coherence to quantum discord}
The relative entropy of coherence of a quantum state $\rho$ is given by $C^r(\rho)=\min_{\sigma\in\mathcal{I}}S(\rho||\sigma)=S(\Delta(\rho))-S(\rho)$, where $S(\rho||\sigma)=\mathrm{Tr}(\rho\log_2\rho)-\mathrm{Tr}(\rho\log_2\sigma)$
is the quantum relative entropy and $\Delta(\rho)=\sum_i|i\rangle\langle i|\rho|i\rangle\langle i|$ is the dephased state in reference basis $\{|i\rangle\}$ of the system, $\mathcal{I}$ denotes
the set of all incoherent (diagonal) states.
Consider bipartite systems $A$ and $B$ with basis $\{|i\rangle_A\}$ and $\{|i\rangle_B\}$, respectively.
The $B$-incoherent states with respect to $\{|i\rangle_B\}$, denoted as $\mathcal{I}_{A|B}$, have the form $\sigma_{AB}=\sum_ip_i\sigma_A^i\otimes |i\rangle_B\langle i|$. A map $\Lambda_{A|B}$ which maps $B$-incoherent states to $B$-incoherent ones is called $B$-incoherent operation.
With respect to $B$-incoherent states, the corresponding coherence is defined by $C^r_{A|B}(\rho_{AB})=\min_{\sigma_{AB}\in\mathcal{I}_{A|B}}S(\rho_{AB}||\sigma_{A|B})
=S(\Delta_B(\rho_{AB}))-S(\rho_{AB})$, where $\Delta_B(\rho_{AB})=\sum_i(\mathbb{I}\otimes|i\rangle_B\langle i|)\rho(\mathbb{I}\otimes|i\rangle_B\langle i|)$ is the local dephasing associated with the subsystem $B$, $\mathbb{I}$ is the identity operator.
Since the relative entropy does not increase under quantum operations, $C^r_{A|B}(\rho_{AB})$ is monotonically nonincreasing under local quantum-incoherent operations and classical communication (LQICC).

With respect to the dephasing on subsystem $B$, the relative entropy of discord for bipartite states $\rho_{AB}$ is given by \cite{mk}, $D^r_{A|B}(\rho_{AB})=\min_{\delta_{AB}\in\mathcal{F}_{A|B}}S(\rho_{AB}||\delta_{A|B})$, where $\mathcal{F}_{A|B}=\sum_{i}p_i\mathcal{F}_A^i\otimes |i\rangle_B\langle i|$ is the set of quantum-classical correlated states. A symmetric version of quantum discord with respect to both dephasing on subsystems $A$ and $B$ is defined by $D^s_{AB}(\rho_{AB})=\min_{\chi_{AB}\in C^c_{AB}}S(\rho_{AB}||\chi_{AB})$, where $\chi_{AB}=\sum_{jk}p_{jk}|j\rangle_A\langle j|\otimes |k\rangle_B\langle k|$, and $C^c_{AB}$ is the set of classical-classical correlated states.
The global discord \cite{global} for bipartite states $\rho_{AB}$ is defined by, $D^g_{A|B}(\rho_{AB})=\min\limits_{\{\Pi^i_B\}}D^g_{\{\Pi^i_B\}}(\rho_{AB})$, where $D^g_{\{\Pi^i_B\}}(\rho_{AB}) = S(\rho_{AB}||\Pi^i_B(\rho_{AB}))-S(\rho_{B}||\Pi^i_B(\rho_{B}))$, $\Pi_B=\{\Pi^i_B\}$ is a complete projective measurement on subsystem $B$, see also the original definition of discord \cite{OZ,HV}.

It is evident from the above definitions that $D^s(\rho_{AB})\leq C^r(\rho_{AB})$ \cite{yys},
as this measure of discord is the minimum amount of the coherence in any product basis \cite{modi}.
We have the following conclusion, see proof in Appendix.

{\bf Theorem 1.} For any bipartite state $\rho_{AB}$, it holds
$D^g_{A|B}( \rho_{AB})+P_{\rho_B}\leq D^r_{A|B}( \rho_{AB})\leq C^r_{A|B}( \rho_{AB})\leq C^r(\rho_{AB})-C^r(\rho_A)$, where $P_{\rho_B}=\min_{\Pi_{B}}S[\pi_{\Pi_{B}(\rho_{AB})}]-S(\pi_{\rho_{AB}})$ with $\pi_{\rho_{AB}}=Tr_B\rho_{AB}\otimes Tr_A\rho_{AB}$ the product of the reduced states.

If the project measurement $\Pi_{B}$ on subsystem $B$ is given by the reference basis $\{|i\rangle_B\}$ of the coherence for subsystem $B$, one can easily get that $P_{\rho_B}=C^r(\rho_B)$ for relative entropy of coherence. Thus Theorem 1 shows that the summation of the global discord with local measurements on subsystem $B$ and the coherence of subsystem $B$ is bounded by $D^r_{A|B}(\rho_{AB})$ and
$C^r_{A|B}(\rho_{AB})$. On the other hand, the $B$-incoherent state of $\rho_{AB}$, $C^r_{A|B}(\rho_{AB})$ (or the discord with local measurements on subsystem $B$, $D^r_{A|B}(\rho_{AB})$), and the coherence of subsystem $A$ is bounded by the coherence $C^r(\rho_{AB})$ of the $\rho_{AB}$. The first two equalities in Theorem 1 hold for some optimal bases $\{|i\rangle_B^*\langle i|\}$ which give the minimum solution of quantum discord $D^r_{A|B}(\rho_B)$. Moreover, if one performs local measurements on subsystem $A$, similar relation can be obtained, $D^g_{B|A}( \rho_{AB})+P_{\rho_A}\leq D^r_{B|A}( \rho_{AB})\leq C^{r_B}_{B|A}( \rho_{AB})\leq C^r(\rho_{AB})-C^r(\rho_B)$.

To illustrate the inequality presented in Theorem 1, let us consider two simple examples. The first one is a two-qubit separable state \cite{xzj,phd}: $\rho_{AB}=\frac{1}{4}[|+\rangle\langle+|\otimes|0\rangle\langle0|
+|-\rangle\langle-|\otimes|1\rangle\langle1|+|0\rangle\langle0|\otimes|
-\rangle\langle-|+|1\rangle\langle1|\otimes|+\rangle\langle+|]$, where $|+\rangle=\frac{1}{\sqrt{2}}(|0\rangle+|1\rangle)$ and $|-\rangle=\frac{1}{\sqrt{2}}(|0\rangle-|1\rangle)$. The optimal basis $\{|i\rangle_B^*\langle i|\}$ which gives the minimum solution of quantum discord $D^r_{A|B}(\rho_B)$ is just $\{|i\rangle_B\langle i|\}$.
Under this basis we have $P_{\rho_B}=0$, $C^r(\rho_A)=C^r(\rho_B)=0$,
$D^g_{A|B}( \rho_{AB})=D^r_{A|B}( \rho_{AB})=C^r_{A|B}( \rho_{AB})\approx 0.311$, and $C^r(\rho_{AB})=0.5$. The first two inequalities in Theorem 1 are equalities in this case.
The second one is the Werner state: $\rho_{AB}=(1-p)\frac{I}{4}+p|\psi\rangle\langle\psi|$, where $|\psi\rangle=\frac{1}{\sqrt{2}}(|00\rangle+|11\rangle)$ is a Bell state, $p\in [0,1]$. The state is a separable for $0<p\leq\frac{1}{3}$ with nonzero discord. The nearest classical state is just the closet incoherent state of $\rho_{AB}$ \cite{luosl}. Under optimal basis $\{|i\rangle_B\langle i|\}$
we have $P_{\rho_B}=0$, $C^r(\rho_A)=C^r(\rho_B)=0$, $D^g_{A|B}( \rho_{AB})=D^r_{A|B}( \rho_{AB})=C^r_{A|B}( \rho_{AB})=C^r(\rho_{AB})$. In this case, all the inequalities in Theorem 1 become
equalities.

The total correlation between systems $A$ and $B$ in a bipartite state $\rho_{AB}$ is given by the quantum mutual information $I(\rho_{AB})=S(\rho_A)+S(\rho_B)-S(\rho_{AB})$. In the following, we show that
the total correlation present in a bipartite state $\rho_{AB}$ is bounded, see proof in Appendix.

{\bf Theorem 2.} For any bipartite state $\rho_{AB}$, we have $I(\rho_{AB})\leq I_T(\rho_{AB})+D^{r_T}_{\bar{T}|T}(\rho_{AB})-P_{\rho_T}$, where $P_{\rho_T}=\min_{\Pi_{T}}S[\pi_{\Pi_{T}(\rho_{AB})}]-S(\pi_{\rho_{AB}})$, $I_T(\rho_{AB})=\max_{\{\Pi^i_T\}}I(\Pi^i_T\rho_{AB}\Pi^i_T)$, $T=A,B,AB$, $\bar{T}$ is the complementary of $T$ in the subsystem of $AB$, with $D^{r}_{\bar{AB}|AB}(\rho_{AB})=D^{s}_{AB}(\rho_{AB})$.

The equality in Theorem 2 holds,
$I(\rho_{AB})= I_{\Pi_T}(\rho_{AB})+D^{r_T}_{\bar{T}|{\Pi_T}}(\rho_{AB})-P_{\rho_{\Pi_T}}$, if the measurement $\Pi_{T}$ on system $T$ is just the reference basis of coherence for $T$, $T=A,B,AB$. When $T=AB$, one gets $I(\rho_{AB})+C^r(\rho_{A})+C^r(\rho_{B})=C^c(\rho_{AB})+D^s_{AB}(\rho_{AB})$, where $C^c(\rho_{AB})=I_{AB}(\rho_{AB})$ \cite{mk} is classical correlation given by the minimal distance between $\rho_{AB}$ and product states $\pi$, $C^c(\rho_{AB})=\min_{\pi} S(\rho_{AB}||\pi)$, with $\rho_{AB}\in C^c_{AB}$.
This means that the sum of the mutual information and the local coherence is equal to the sum of the quantum discord and classical correlations.
One can also obtain that $I(\rho_{AB})-C^c(\rho_{AB})=D^s_{AB}(\rho_{AB})-C^r(\rho_{A})-C^r(\rho_{B})$, which means that the overall quantum correlations given in a bipartite state $\rho_{AB}$ is equal to the quantum discord minus the coherence in each subsystem. When $T=A(B)$, one obtains $I(\rho_{AB})+C^r(\rho_B)= I_B(\rho_{AB})+D^r_{A|B}(\rho_{AB})$, namely, the sum of the mutual information and the coherence of the measured subsystem $B(A)$ is equal to the sum of the discord and conditional mutual information performed on subsystem $B(A)$.

{\sf Example 1} Let us consider the Bell-diagonal states \cite{rmhoro,hjx},
$\rho_{AB}=\frac{1}{4}(I\otimes I+\sum_{j=1}^3c_j\sigma_j\otimes\sigma_j)$,
where $\sigma_j$ are the standard Pauli matrices. In this case, we have $I(\rho_{AB})-I_{AB}(\rho_{AB})=I(\rho_{AB})- C^c(\rho_{AB})=D^g_{AB}(\rho_{AB})={D}^s_{AB}(\rho_{AB})$, and $I(\rho_{AB})-I_B(\rho_{AB})=D^g_{A|B}(\rho_{AB})=D^r_{A|B}(\rho_{AB})$, see Appendix for detailed derivations.

\section{Discord distribution between spatially separated parties}
Consider two remote agents, Alice and Bob, having access to particles $A$ and $B$, respectively.
Alice interacts an ancilla $C$ with her particle $A$ and sends $C$ to Bob. Bob interacts $C$ with his particle $B$. At the end how much discord they share could be increased? What is
the cost to increase the discord they share by sending an auxiliary quantum particle $C$? see Fig 1.
\begin{figure}
  \centering
  \includegraphics[width=8cm]{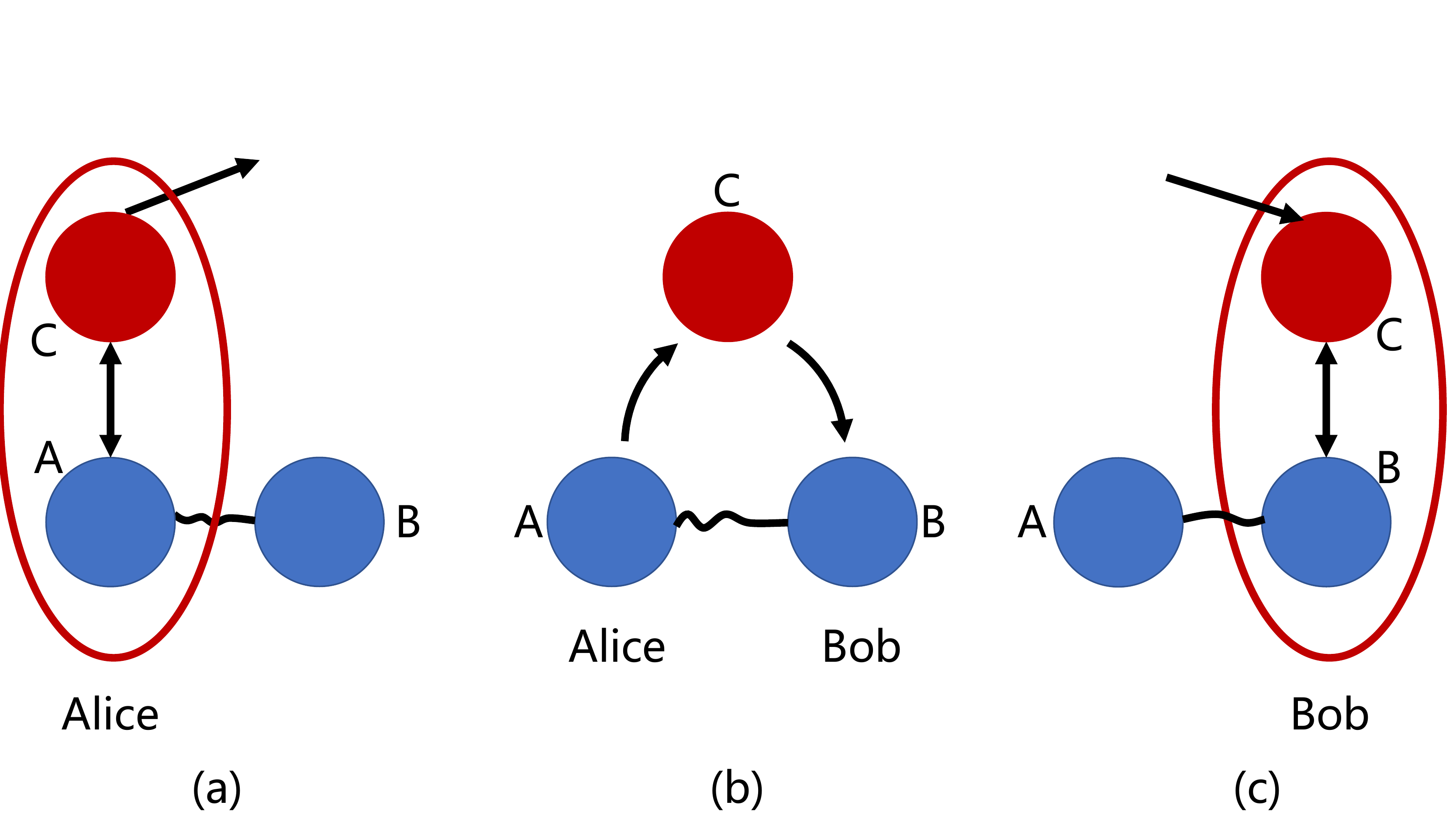}\\
  \caption{Alice and Bob have particles $A$ and $B$, respectively. (a) Alice interacts an ancilla $C$ with her particle $A$. (b) $C$ is then sent to Bob's side. (c) Bob interacts $C$ with his particle $B$.}
\end{figure}

Let $\rho$ be the initial state of the particles $A$, $B$ and $C$. The initial discord between Alice and Bob is $D^{r}_{AC|B}(\rho)$. If the particle $C$ is sent to Bob's side without any operations, the discord between the them is given by $D^{r}_{A|BC}(\rho)$. We first present a general relation among
$D^r_{AC|B}(\rho)$, $D^r_{A|BC}(\rho)$ and the cost $C^r_{AB|C}(\rho)$ for arbitrarily given $\rho$.
Consider the optimal projective measurement $\Pi_C^*=\{|i\rangle_C\langle i|\}$ on $C$, with $p_i$ the probability of outcome $i$ and $\rho_{AB}^i$ the corresponding conditional states of systems $AB$, i.e., $\Pi_C^*(\rho_{ABC})=\sum_i p_i\rho_{AB}^i\otimes |i\rangle_C\langle i|$. Then we have the following result, see proof in Appendix.

{\it Theorem 3.}--For any tripartite state $\rho$ of systems $A$, $B$ and $C$, it holds that
\begin{eqnarray}\label{th1}
D^r_{T|\bar{T}C}(\rho)-D^r_{TC|\bar{T}}(\rho)\leq C^r_{AB|C}(\rho),
\end{eqnarray}
where $T=A,B$, and $\bar{T}$ is the complementary of $T$ in the subsystem $AB$.

We point out that the inequality (\ref{th1}) holds for any dimensions of the subsystems. The implications of Theorem 1 is illustrated in Fig. 2.
In particular, for tripartite pure state $\rho_{ABC}=|\psi\rangle_{ABC}\langle\psi|$
from Theorem 3 we have
\begin{eqnarray*}
|S(\rho_{A})-S(\rho_{B})|\leq S[\Delta(\rho_{AB})],
\end{eqnarray*}
where $\Delta$ is a full dephasing operation.
 \begin{figure}
  \centering
  \includegraphics[width=10cm]{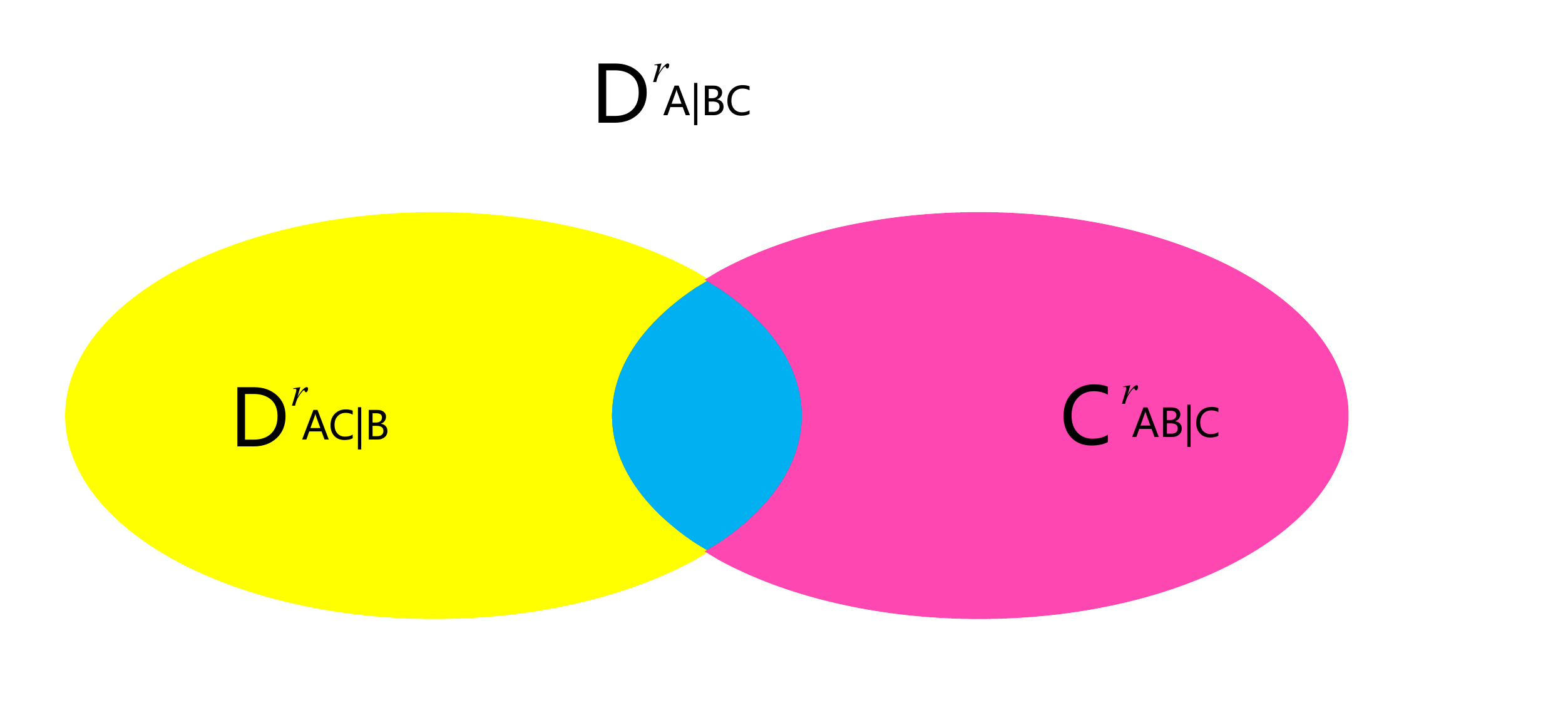}\\
  \caption{The area (yellow and blue) represents the discord between $AC$ and $B$, while the area (red and blue) represents the quantum correlations between $AB$ and $C$. The total area (yellow, blue and red) represents the discord between $A$ and $BC$. One can read off the main result: $D^r_{A|BC}(\rho)-D^r_{AC|B}(\rho)\leq C^r_{AB|C}(\rho)$.}
\end{figure}

Now let $\alpha$ denote the initial state of $A$, $B$ and $C$, and $\beta$ the state after Alice interacts the ancilla $C$ with particle $A$. As local operation on $AC$ cannot increase the discord in the $AC|B$ cut, one has $D^r_{AC|B}(\beta)\leq D^r_{AC|B}(\alpha)$. Then Alice sends $C$ to Bob, who interacts $C$ with particle $B$. From Theorem 3 for state $\beta$ one gets $D^r_{A|BC}(\beta)\leq D^r_{AC|B}(\alpha)+C^r_{AB|C}(\beta)$. This shows that the discord gained between Alice and Bob is bounded by the quantum coherence measured on $C$. 

It is impossible to distribute the discord by LQICC. Let us first address the case of $C^r_{AB|C}(\rho)=0$, i.e., $\rho$ is a quantum-incoherent state, $\rho=\sum_i p_i \rho^{i}_{AB}\otimes |i\rangle_C\langle i|$, which corresponds to classical communication from Alice to Bob. The index $i$ embodies classical information that Alice may copy locally before sending $C$ to Bob. Then both Alice and Bob have access to this information after $C$ is transferred from Alice to Bob, and a local incoherent transformation can be performed by Bob depending on the index $i$.
The process is just the one communication step for a general protocol in terms of LQICC. 
The protocol may also include the round of classical communication with $C$ that is sent from Bob to Alice. Then one obtains $D^r_{B|AC}(\beta)\leq D^r_{BC|A}(\alpha)+ C^r_{AB|C}(\beta).$
In this case, local classical registers can be kept or erased at any stage of the protocol. Inequality (\ref{th1}) gives rise to that coherence does not increase at any step of a protocol based on LQICC. If $C^r_{AB|C}(\rho)$ does not vanish, the transfer of $C$ cannot be interpreted as classical communication, revealing the role of coherence in general quantum communication. Hence, (\ref{th1}) constitutes a nontrivial relaxation of the condition of monotonicity of discord under LQICC, bounding the increase of discord under local quantum-incoherent operations and quantum communication.

In order to investigate the discord distribution via a quantum-classical system, besides the coherence present in $\beta$, there must be coherence on the receiver's side already in the initial state $\alpha$. Exchanging the roles of $B$ and $C$, one gets from (\ref{th1}), $D^r_{A|BC}(\beta)-D^r_{AB|C}(\beta)\leq C^r_{AC|B}(\alpha)$. Suppose $C$ is a classical state, i.e., $D^r_{AB|C}(\beta)=0$, we obtain the relation $D^r_{A|BC}(\beta)\leq C^r_{AC|B}(\alpha)$. Note that if $C$ is initially not correlated with $AB$, one further gets $D^r_{A|BC}(\beta)\leq C^r_{A|B}(\alpha)$. Another interesting case $C^r_{AC|B}(\alpha)=0$. Then $B$ is incoherent state initially, and hence $\beta=\sum_ip_i\beta^i_{AC}\otimes|i\rangle_B\langle i|$. In this case discord between Alice and Bob can only be created if $C$ and $A$ ($B$) have non-vanishing discord, in particular, only if at least one $\beta^i_{AC}$ has non-vanishing discord. Indeed, such $\beta$ simply describes a situation in which Bob, upon reading the index $i$ interacted in B, knows which states $\beta^i_{AC}$ he will end up sharing with Alice. Let us consider two examples.

{\sf Example 2} Discord distribution with non-vanishing initial discord between Alice and Bob. Let us consider the state $\rho=|+\rangle_A\langle+|\otimes|0\rangle_C\langle0|\otimes|-\rangle_B\langle-|$, where $|+\rangle_A=\frac{1}{\sqrt{2}}(|0\rangle+|1\rangle)$ and $|-\rangle_B=\frac{1}{\sqrt{2}}(|0\rangle-|1\rangle)$. Alice applies an incoherent operation $\mathcal{O}(\rho_{ACB})=pU_{AC}\rho_{ACB} U_{AC}^\dagger+(1-p)\frac{I}{4}\otimes|-\rangle_B\langle-|$, where $I$ is unit operator, $0\leq p\leq1$, $U_{AC}$ is the CNOT gate $U_{AC}(|i\rangle\otimes|j\rangle)=|i\rangle\otimes|i\oplus j\rangle$. The output state is $\rho_1=p|\Psi\rangle\langle\Psi|\otimes|-\rangle_B\langle-|+(1-p)\frac{I}{4}\otimes|-\rangle_B\langle-|$, with $|\Psi\rangle=\frac{1}{\sqrt{2}}(|00\rangle+|11\rangle)$. Obviously, the discord between subsystems $A$ and $C$ is greater than 0 for $p>0$, and the entanglement is vanished when $p\leq\frac{1}{3}$. From the inequality (\ref{th1}), we have that the discord between $A$ and $BC$ is bounded by $C^r_{AB|C}(\rho_1)$ after $A$ interacts with $C$, i.e., $D^r_{A|BC}(\rho_1)- D^r_{AC|B}(\rho)\leq \frac{1-p}{4}\log\frac{1-p}{4}+\frac{1+3p}{4}\log\frac{1+3p}{4}-\frac{1+p}{2}\log\frac{1+p}{4}$.

{\sf Example 3} Discord distribution with vanishing initial discord between Alice and Bob.
Consider the initial three-qubit state in Ref. \cite{tsc},
$\alpha = \left( \frac{1}{3}|\phi^{+}\rangle \langle\phi^{+}|+\frac{1}{6}|01\rangle\langle01|+\frac{1}{6}|10\rangle \langle10| \right) \otimes |0\rangle_B \langle0| + \left( \frac{1}{6}|00\rangle\langle00|+\frac{1}{6}|11\rangle \langle11| \right) \otimes |1\rangle_B \langle1|$,
where $|\phi^{+}\rangle = \frac{1}{\sqrt{2}}(|00\rangle+|11\rangle)$ is the maximally entangled state of $A$ and $C$. Alice performs a $\textsc{CNOT}$ operation on $A$ and $C$ with $A$ as the control qubit, and passes $C$ to Bob. Bob performs another $\textsc{CNOT}$ operation on the system $BC$ with $B$ as the control qubit, i.e., $\alpha \xrightarrow{\textsc{CNOT}_{AC}} \beta \xrightarrow{\textsc{CNOT}_{BC}} \gamma$. It shows that the qubit $B$ has zero discord with $A$ and $C$ all the time. Nevertheless, $A$ and $C$ may share some discord at last, $D^r_{A|BC}(\gamma)\leq \frac{1}{3}\log2$.

In fact, one may obtain similar results for other quantum correlations such as information deficit, which quantifies the amount of information that cannot be localized by classical communication between two parties. If only one-way classical communication from party $X$ to party $Y$ is allowed, one has the one-way information deficit:
$\Delta_{X|Y}(\rho_{XY})=\min_{\Pi_Y^i}S(\rho_{XY}||\sum_i\Pi_Y^i\rho_{XY}\Pi_Y^i)$,
where the minimization goes over all local von Neumann measurements $\{\Pi_Y^i\}$ on subsystem $Y$. We have \begin{equation}\label{deficit}
\Delta_{A|BC}(\rho)-\Delta_{AC|B}(\rho)\leq C^r_{AB|C}(\rho),
\end{equation}
see proof in Appendix. 
This shows that the deficit of the bipartite partition $A|BC$ cannot be larger than the sum of the deficit of the partition $AC|B$ plus the quantum coherence across the partition $AB|C$. Thus, the inequality (\ref{deficit}) may be viewed as a type of monogamy relation satisfied by a tripartite quantum state.

\section{conclusion}
Establishing quantum correlations between two distant parties is an essential step of many protocols in quantum information processing. The purpose of the physical transmission of the carrier system is to change the amount of quantum correlations between the between remote agents. For example, the increase of the total correlations, mutual information, is bounded by the amount of communicated correlations \cite{nielsen}, i.e. $I(\rho_{T|\bar{T}C})-I(\rho_{TC|\bar{T}})\leq I(\rho_{AC})\leq I(\rho_{T\bar{T}|C})$, with $T=A,B$, and $\bar{T}$ the complementary of $T$ in the subsystem $AB$.
We have investigated the trade-off relations satisfied by discord and coherence during such essential steps, via distributing the coherence in a bipartite quantum state to the discord between the subsystems, based on the relations between relative entropy and mutual information. We have identified quantum correlations as the key resource for discord distribution and derived a general bound on the discord gained between distant paries under local quantum-incoherent operations and quantum communication. Explicitly, we have proved that the discord gained between distant parties is bounded by the amount of quantum coherence between them and the information carrier, which provides a fundamental connection between quantum discord and quantum coherence and a natural operational interpretation of quantum coherence as the necessary prerequisite for the success of discord distribution. Our results may highlight further studies on quantum resources consuming in information processing and give rise to related experimental demonstrations.

\bigskip

\noindent{\bf Acknowledgments}\, \,
This work was supported in part by the National Natural Science Foundation of China(NSFC) under Grants 11847209; 11675113; 11975236 and 11635009; Beijing Municipal Commission of Education (KZ201810028042); Beijing Natural Science Foundation (Grant No. Z190005; Academy for Multidisciplinary Studies, Capital Normal University; Shenzhen Institute for Quantum Science and Engineering, Southern University of Science and Technology, Shenzhen 518055, China (No. SIQSE202001); the China Postdoctoral Science Foundation funded project No. 2019M650811 and the China Scholarship Council No. 201904910005.

\bigskip
\section*{APPENDIX}
\setcounter{equation}{0} \renewcommand%
\theequation{A\arabic{equation}}

\subsection{Proof of Theorem 1}
It can be shown that $D^r_{A|B}(\rho_{AB})$ corresponds to the minimal entropic increase resulting from the complete projective measurement $\Pi_B$ on $B$: $D^r_{A|B}(\rho_{AB})=\min_{\Pi_B}S[\Pi_B(\rho_{AB})]-S(\rho_{AB})$, where $\Pi_B(\rho_{AB})$ is the state after the measurement $\Pi_B$, $\Pi _{B}( \rho_{AB})=\sum_i(\mathbb{I}\otimes\Pi^i_B)\rho_{AB}(\mathbb{I}\otimes\Pi^i_B)=\sum_i p_i\rho_A^i\otimes |i\rangle_B\langle i|$. Similarly, $D^s_{AB}(\rho_{AB})=\min_{\Pi_{AB}}S[\Pi_{AB}(\rho_{AB})]-S(\rho_{AB})$, where $\Pi_{AB}(\rho_{AB})$ is the state after the measurement $\Pi_{AB}$, $\Pi _{AB}\left( \rho_{AB}\right) =\sum_{j,k} \left(\Pi_{A}^{j}\otimes \Pi_{B}^{k} \right) \rho_{AB} \left(\Pi_{A}^{j}\otimes \Pi_{B}^{k}\right)=\sum_{jk}p_{jk}|j\rangle_A\langle j|\otimes |k\rangle_B\langle k|$. Note that $C^r_{A|B}(\rho_{AB})$ is different from the relative entropy of discord which involves a minimization over all bases of $B$, while $C^r_{A|B}(\rho_{AB})$ is defined for a fixed incoherent basis $\{|i\rangle_B\}$.

Under the von Neumann projective measurement $\Pi_B=\{\Pi^i_B\}$, the state of the system $B$ is given by $\rho_A^i=\mathrm{Tr}_B[(\mathbb{I}\otimes \Pi^i_B)\rho_{AB}(\mathbb{I}\otimes \Pi^i_B)]/p_i$, with the measurement outcome probability $p_i=\mathrm{Tr}[(\mathbb{I}\otimes \Pi^i_B)\rho_{AB}]$. The conditional entropy of system $A$ is then $S[\Pi_B(\rho_{A|B})]=\sum_ip_i S(\rho_A^i)$. Therefore, the quantum mutual information induced by the von Neumann measurement on the system $B$ is given by
\begin{eqnarray*}
I_{\Pi_B}(\rho_{AB})=S(\rho_A)-S[\Pi_B(\rho_{A|B})].
\end{eqnarray*}
The measurement-independent quantum mutual information $I_B(\rho_{AB})$ is given by
\begin{eqnarray*}
I_B(\rho_{AB})&&=\max_{\Pi_B}I_{\Pi_B}(\rho_{AB})\nonumber\\
&&=\max_{\{\Pi^i_B\}}I\left((\mathbb{I}\otimes\Pi^i_B)\rho_{AB}(\mathbb{I}\otimes\Pi^i_B)\right),
\end{eqnarray*}
which is interpreted as the one-sided classical mutual information on subsystem $B$.
Let $\Pi_B^*=\{|i\rangle_B^*\langle i|\}$ be the optimal basis of system $B$ for $D^r_{A|B}$. Then we have \begin{eqnarray*}
&&D^g_{A|B}(\rho_{AB})\nonumber\\&&=\min_{\Pi_B}S(\rho_B)-S(\rho_{AB})+\sum_ip_iS(\rho_A^i)\nonumber\\
&&\leq S(\rho_B)-S(\rho_{AB})+S[\Pi^* _{B}(\rho_{AB})]-S[\Pi^*_{B}( \rho_{B})]\nonumber\\
&&=S(\rho_B)+S[\mathrm{tr}_B\Pi^*_{B}(\rho_{AB})]-S(\rho_{AB})+S([\Pi^*_{B}(\rho_{AB})]\nonumber\\
&&~~~-S[\mathrm{tr}_A\Pi^*_{B}(\rho_{AB})]-S[\mathrm{tr}_B\Pi^*_{B}(\rho_{AB})]\nonumber\\
&&=S(\pi_\rho)-S(\rho_{AB})+S[\Pi^*_{B}(\rho_{AB})]-S(\pi_{\Pi^*_{B}(\rho_{AB})})\nonumber\\
&&=D^r_{A|B}( \rho_{AB})-[S(\pi_{\Pi^* _{B}(\rho_{AB})})-S(\pi_\rho)]\nonumber\\
&&\leq D^r_{A|B}( \rho_{AB})-P_{\rho_B},
\end{eqnarray*}
where we have used $S[\mathrm{tr}_A\Pi^* _{B}(\rho_{AB})]=S[\Pi^*_{B}( \rho_{B})]$ and $\Pi^*_{B}( \rho_{B})=\sum_ip_i |i\rangle^*_B\langle i|$.
Then one gets
\begin{eqnarray*}
D^g_{A|B}( \rho_{AB})+P_{\rho_B}\leq D^r_{A|B}( \rho_{AB}).
\end{eqnarray*}
It is evident that $D^r_{A|B}( \rho_{AB})\leq C^r_{A|B}( \rho_{AB})$ for any reference basis \cite{yys}. In fact, the measure of discord is the minimum coherence in any product basis \cite{modi}.
The equality holds for an optimal basis. 
On the other hand,
\begin{eqnarray}
C^r_{A|B}( \rho_{AB})&&=S[\Delta_B(\rho_{AB})]-S(\rho_{AB})\nonumber\\
&&=S[\Delta_{AB}(\rho_{AB})]-S(\rho_{AB})\nonumber\\
&&-\left(S[\Delta_{AB}(\rho_{AB})]-S[\Delta_B(\rho_{AB})]\right)\nonumber\\
&&\leq C^r(\rho_{AB})-C^r(\rho_A),
\end{eqnarray}
where $\Delta_{AB}$ is the completely dephasing operation.

\subsection{Proof of Theorem 2}
The total mutual information of a bipartite state $\rho$ is given by the relative entropy between $\rho$
the product state of the reduced states $\pi_\rho=\rho_A\otimes\rho_B$,
$I(\rho_{AB})=S(\rho_{AB}||\pi_{\rho_{AB}})=S(\rho_{AB}||\rho_A\otimes\rho_B$) \cite{mk}. We have
\begin{eqnarray}
I_B(\rho_{AB})&&=S(\rho_A)-\sum_ip_iS(\rho_A^i)\nonumber\\
&&=S(\rho_A)+\sum_iS(\Pi^i_B(\rho_{B}))\nonumber\\
&&~~~-[\sum_iS(\Pi^i_B(\rho_{B}))+\sum_ip_iS(\rho_A^i)]\nonumber\\
&&=S(\rho_A)+\sum_iS(\Pi^i_B(\rho_{B}))-\sum_iS(\Pi^i_B(\rho_{AB}))\nonumber\\
&&=S\left( \Pi _{B}( \rho_{AB}) \parallel \rho_{A}\otimes \Pi _{B}( \rho_{B}) \right)\nonumber\\
&&=S(\rho_A)+S(\Pi _{B}( \rho_{B}))-S(\Pi _{B}( \rho_{AB})),
\end{eqnarray}
with maximization taken over measurement $\{\Pi_{B}^{j}\}$, where $\Pi _{B}( \rho_{AB})=\sum_i(\mathbb{I}\otimes\Pi^i_B)\rho_{AB}(\mathbb{I}\otimes\Pi^i_B)=\sum_i p_i\rho_A^i\otimes |i\rangle_B\langle i|$ and $\Pi _{B}( \rho_{B})=\sum_ip_i |i\rangle_B\langle i|$.

Combining
\begin{eqnarray*}
D^r_{A|B}(\rho_{AB})=\min_{\Pi_B}S[\Pi_B(\rho_{AB})]-S(\rho_{AB})
\end{eqnarray*}
and
\begin{eqnarray*}
P_{\rho_B}&&=\min_{\Pi_B}S[\pi_{\Pi_{B}(\rho_{AB})}]-S(\pi_{\rho_{AB}})\nonumber\\
&&=\min_{\Pi_B}S(\Pi _{B}( \rho_{B}))-S(\rho_B)
\end{eqnarray*}
we have $I(\rho_{AB})= I_{\Pi_T}(\rho_{AB})+D^r_{\bar{T}|{\Pi_T}}(\rho_{AB})-P_{\rho_{\Pi_T}}$, where $I_{\Pi_T}(\rho_{AB})$, $D^r_{\bar{T}|{\Pi_T}}(\rho_{AB})$ and $P_{\rho_{\Pi_T}}$ are the projective measurement $\Pi_T$ dependent, $T=A,B,AB$, and $\bar{T}$ is the complementary of $T$ in the subsystem $AB$. Under the optimal local measurements, one has $I(\rho_{AB})\leq I_B(\rho_{AB})+D^r_{A|B}(\rho_{AB})-P_{\rho_B}$.
With a similar consideration, we can also get $I(\rho_{AB})\leq I_A(\rho_{AB})+D^r_{B|A}(\rho_{AB})-P_{\rho_A}$ and $I(\rho_{AB})\leq I_{AB}(\rho_{AB})+D^s_{AB}(\rho_{AB})-P_{\rho_{AB}}$.

\subsection{Derivations in Example 1}
Consider the Bell-diagonal states \cite{rmhoro,hjx},
$\rho_{AB}=\frac{1}{4}(I\otimes I+\sum_{j=1}^3c_j\sigma_j\otimes\sigma_j)=\sum_{ab}\lambda_{ab}|\beta_{ab}\rangle\langle\beta_{ab}|$,
with the maximally mixed marginals ($\rho_A=\rho_B=\frac{I}{2}$).
The density matrix of Bell-diagonal states with $\sigma_3$ representation takes the form
\begin{equation*}\label{}
  \rho_{AB}^{\sigma_3}=\frac{1}{4}\left(
    \begin{array}{cccc}
      1+c_3 & 0 & 0 & c_1-c_2 \\
      0 & 1-c_3 & c_1+c_2 & 0 \\
      0 & c_1+c_2 & 1-c_3 & 0 \\
      c_1-c_2 & 0 & 0 & 1+c_3 
    \end{array}
  \right).
\end{equation*}
The eigenstates of $\rho_{AB}^{\sigma_3}$ are the four Bell states:
$|\beta_{ab}\rangle=(|{0,b}\rangle+(-1)^a|{1,1\oplus b}\rangle)/\sqrt{2}$,
with the corresponding eigenvalues
$\lambda_{ab}=\frac{1}{4}[1+(-1)^ac_1-(-1)^{a+b}c_2+(-1)^bc_3]$,
where $a,b\in\{0,1\}$.

For Bell-diagonal states, the reduced states have no coherence in the subsystems.
The relative entropy of coherence is given by
\begin{eqnarray*}\label{co}
C^r(\rho_{AB}^{\sigma_i}) =-H(\lambda_{ab})-\sum_{j=1}^2\frac{(1+(-1)^jc_i)}{2}\log_2\frac{(1+(-1)^jc_i)}{4},
\end{eqnarray*}
where $H(\lambda_{ab})=-\sum_{ab}\lambda_{a,b}\log_2\lambda_{ab}$.
The mutual information for Bell-diagonal states is given by
\begin{eqnarray*}
I(\rho_{AB})=\sum_{a,b}\lambda_{ab}\log_2(4\lambda_{ab}).
\end{eqnarray*}
The classical correlation for Bell-diagonal states is given by
\begin{eqnarray*}
C^c(\rho_{AB})=\sum_{j=1}^2\frac{(1+(-1)^jc)}{2}\log_2(1+(-1)^jc),
\end{eqnarray*}
where $c=\max\{|c_1|,|c_2|,|c_3|\}$.

Before calculating $D^g_{AB}(\rho_{AB})$, we note that
from the derivation of Theorem 2, the quantum discord can be rewritten as the difference of relative entropies:
\begin{eqnarray*}
D^g_{A|B}( \rho_{AB})
&=&I(\rho_{AB})-I_B(\rho_{AB})\nonumber\\
&=& S\left(\rho_{AB}\parallel \rho_{A}\otimes \rho_{B}\right)\nonumber\\
&&-S( \Pi _{B}\left( \rho_{AB}) \parallel \rho_{A}\otimes \Pi _{B}( \rho
_{B}) \right)\nonumber\\
&=&S[\rho_{AB}||\Pi_B( \rho_{AB})]-S[\rho_{B}||\Pi_B( \rho_{B})],
\end{eqnarray*}
with the minimization taken over the measurement $\{\Pi_{B}^{j}\}$.
Performing measurements on both subsystems $A$ and $B$, one has the symmetric version $D^g_{AB}\left(\rho_{AB}\right)$,
\begin{eqnarray}\label{discord}
D^g_{AB}\left(\rho_{AB}\right) &=& \min_{\{\Pi_{A}^{j}\otimes\Pi_{B}^{k}\}}
\left[  S\left( \rho_{AB}\parallel \rho_{A}\otimes \rho_{B}\right) \right. \nonumber \\
&&\left.\hspace{-0.6cm}-S\left( \Pi _{AB}\left(\rho_{AB}\right) \parallel \Pi _{A}\left( \rho_{A}\right) \otimes \Pi
_{B}\left(\rho_{B}\right) \right) \right],
\end{eqnarray}
where $\Pi _{AB}\left( \rho_{AB}\right) =\sum_{j,k} \left(\Pi_{A}^{j}\otimes
\Pi_{B}^{k} \right) \rho_{AB} \left(\Pi_{A}^{j}\otimes \Pi_{B}^{k}\right)$.
Expressing (\ref{discord}) in terms of the mutual information $I$, we obtain
\begin{eqnarray}\label{di}
D^g_{AB}\left( \rho_{AB}\right) =
\min_{\{\Pi_{A}^{j}\otimes\Pi_{B}^{k}\}} \left[
I(\rho_{AB}) -
I(\Pi _{AB}\left( \rho_{AB}\right))\right],
\end{eqnarray}
which is the symmetric version of the
expression for the loss of correlation based on the measurement~\cite{Luo:10,Okrasa:11}.
Remarkably, $D^g_{AB}\left( \rho_{AB}\right)$ is equivalent to the
measurement-induced disturbance \cite{Luo:08} if the measurements performed (\ref{discord})
are replaced by the eigenprojectors of the reduced density operators, respectively. Moreover, Eq.~(\ref{discord}) also provides the symmetric quantum discord considered in Ref.~\cite{Maziero:10} and experimentally witnessed in Ref.~\cite{Auccaise:11}.
Eq.~(\ref{discord}) yields
\begin{eqnarray*}
D^g_{AB}\left( \rho_{AB}\right) &=& \min_{\{\Pi_{A}^{j}\otimes\Pi_{B}^{k}\}}
\left[ S\left( \rho_{AB}\parallel \Pi
_{AB}\left( \rho_{AB}\right) \right) \right. \nonumber \\
&&\left.\hspace{-0.6cm}-S\left( \rho_{A}\parallel
\Pi _{A}\left( \rho_{A}\right) \right) -S\left( \rho_{B}\parallel \Pi _{B}\left( \rho_{B}\right) \right)\right] \nonumber \\
&&\leq  S\left( \rho_{AB}\parallel \Pi
_{AB}\left( \rho_{AB}\right) \right)-P_{\rho_A}-P_{\rho_B}.
\end{eqnarray*}
Specially, for some basis the symmetric extension quantum discord $D^g_{AB}\left(\rho_{AB}\right)$ is bounded by the correlated coherence $C_{cc}(\rho_{AB})=C^r(\rho_{AB})-C^r(\rho_{A})-C^r(\rho_{B})$ defined in \cite{kct}.

From (\ref{di}), we have the quantum discord for Bell-diagonal states,
\begin{eqnarray*}
D^g_{AB}(\rho_{AB})
=-H(\lambda_{ab})-\sum_{j=1}^2\frac{(1+(-1)^jc)}{2}\log_2\frac{(1+(-1)^jc)}{4}.\end{eqnarray*}

We note that the one-side quantum discord, two-side quantum discord and the relative entropy of quantum discord are identical for Bell-diagonal states. It is easy to verify that the quantum discord is equal to the coherence under an optimal basis. Therefore, $I(\rho_{AB})-I_B(\rho_{AB})=D^g_{A|B}(\rho_{AB})=D^r_{A|B}(\rho_{AB})$ as $C^r(\rho_A)=C^r(\rho_B)=0$ and $I(\rho_{AB})-I_{AB}(\rho_{AB})=I(\rho_{AB})- C^c(\rho_{AB})=D^g_{AB}(\rho_{AB})={D}^s_{AB}(\rho_{AB})$.

\subsection{Proof of Theorem 3}
Let $\rho_{AB}^{i*}$ be the state resulted from the optimal measurement on subsystem $B$ for  $D^r_{A|B}(\rho_{AB}^i)$. As the state $\sum_i p_i\rho_{AB}^{i*}\otimes |i\rangle_C\langle i|$ is a quantum-classical state, we have
\begin{eqnarray*}\label{}
&&D^r_{A|BC}(\rho)\leq S(\rho\|\sum_i p_i \rho^{i*}_{AB}\otimes |i\rangle_C\langle i| ) \\
&&=-S(\rho)-\mathrm{Tr}[ \rho \log (\sum_i p_i \rho^{i*}_{AB}\otimes |i\rangle_C\langle i| ) ] \\
&&=-S(\rho)-\mathrm{Tr}[ \Pi^*_C(\rho) \log (\sum_i p_i \rho^{i*}_{AB}\otimes |i\rangle_C\langle i| )] \\
&&=\Big[S(\Pi^*_C(\rho))-S(\rho)\Big]\notag+\Big[-S(\Pi^*_C(\rho))\\
&&\quad-\mathrm{Tr}\Big(\Pi^*_C(\rho) \log ( \sum_i p_i \rho^{i*}_{AB}\otimes |i\rangle_C\langle i|) \Big)\Big]\notag\\
&&=C^r_{AB|C}(\rho)+S(\sum_i p_i \rho^i_{AB}\otimes |i\rangle_C\langle i|\\
&&\quad\|\sum_i p_i \rho^{i*}_{A:B}\otimes |i\rangle_C\langle i|)\\
&&= C^r_{AB|C}(\rho)+\sum_ip_iS(\rho^i_{AB}\| \rho^{i*}_{AB})\\
&&=C^r_{AB|C}(\rho) + \sum_ip_i D^r_{A|B}(\rho^i_{AB})\\
&&=C^r_{AB|C}(\rho) + D^r_{A|BC} (\Pi^*_C(\rho)) \\
&&= C^r_{AB|C}(\rho) + D^r_{AC|B} (\Pi^*_C(\rho))\\
&&\leq C^r_{AB|C}(\rho) + D^r_{AC|B} (\rho),
\end{eqnarray*}
where the first inequality is due to that the quantum-classical state $\sum_i p_i \rho^{i*}_{A:B}\otimes |i\rangle_C\langle i|$ cannot be better than optimal state for the sake of $D^r_{A|BC}(\rho)$, the second equality holds since $\mathrm{Tr}(\sigma\log\Pi(\tau))=\mathrm{Tr}(\Pi(\sigma)\log\Pi(\tau))$ for all projective measurements $\Pi$, and for all $\sigma$ and $\tau$~\cite{ni}, the fourth equality is due to the optimality of $\Pi^*_C$ for $C^r_{AB|C}(\rho)$, the fifth equality is due to the chain rule for relative entropy~\cite{mpi}, the last two equalities are due to the fact that the relative entropy of coherence satisfies the ``flags'' condition Ref.~\cite{mhoro}, i.e., $D^r_{FX|Y}\left(\sum_i p_i |i\rangle_F\langle i|\otimes\rho^i_{XY}\right)=\sum_i p_i D^r_{X|Y}(\rho^i_{XY})= D^r_{X|YF}\left(\sum_i p_i\rho^i_{XY}\otimes |i\rangle_F\langle i|\right)$.
From the above consideration, the cost for sending the particle $C$ from Alice to Bob is bounded by $C^r_{AB|C}(\rho)$, $D^r_{B|AC}(\rho)-D^r_{BC|A}(\rho)\leq C^r_{AB|C}(\rho)$.

{\sf Example.} Let us consider the state $\rho=|+\rangle_A\langle+|\otimes|0\rangle_B\langle0|\otimes|0\rangle_C\langle0|$, where $|+\rangle_A=\frac{1}{\sqrt{2}}(|0\rangle+|1\rangle)$.  Define an incoherent operation as $\varepsilon(\rho)=\frac{1}{2}(U_{CN}\otimes I_B)\rho(U_{CN}^\dagger\otimes I_B)+\frac{1}{8}I$ on subsystem $AC$, where $I_B$ is unit operator on subsystem $B$, $U_{CN}$ is the CNOT gate $U_{CN}(|i\rangle\otimes|j\rangle)=|i\rangle\otimes|i\oplus j\rangle$.
Alice applies the incoherent operation $\varepsilon$ on initial state $\rho$ and passes qubit $C$ to Bob, who then performs yet another CNOT operation $U'_{CN}$ on the subsystem $BC$ with $C$ as the control qubit. 
The resulting state is $\rho'_{ABC}=\frac{3}{8}(|000\rangle\langle 000|+|111\rangle\langle 111|)+\frac{1}{4}|000\rangle\langle 111|+|111\rangle\langle 000|)+\frac{1}{8}(|011\rangle\langle 011|+|100\rangle\langle 100|)$.
Using concurrence $E_C$ as the entanglement measure, we have $E_C(\rho'_{A|BC})\geq E_C(\rho'_{AB})=\frac{1}{4}$ (The concurrence of a two-qubit mixed state $\rho$ is given by $E_C(\rho) = \mathrm{max}\{0,~\lambda_1-\lambda_2 -\lambda_3 -\lambda_4\}$, where $\lambda_1,~\lambda_2,~\lambda_3,~\lambda_4$ are the square root of the eigenvalues of $\rho(\sigma_y \otimes \sigma_y )\rho^{\star}(\sigma_y \otimes \sigma_y )$ in nonincreasing order, $\sigma_y$ is the Pauli matrix, and $\rho^{\star}$ is the complex conjugate of $\rho$).
That is to say, the final state $\rho'_{ABC}$ is entangled. Thus the discord of the final state $\rho'_{ABC}$ is nonzero, i.e.,$D^r(\rho'_{A|BC})>0$, while the discord of the initial state $\rho$ is 0. Therefore, discord must have strictly increased through the transfer of a separable carrier.  From inequality (1), we have the increase of the discord is bounded by the coherence between the carrier and two remote parties $C^r(\rho'_{AB|C})=0.182$, i.e., $D^r(\rho'_{A|BC})\leq C^r(\rho'_{AB|C})=0.182$.

\subsection{Derivation of (\ref{deficit})}
Let $\sigma=\sum_i\Pi_B^i\rho\Pi_B^i$ be the state from the local measurement $\Pi_B=\{\Pi_B^i\}$ on the $B$ part of $\rho$, which minimizes the relative entropy of $\sigma$ and $\rho$ such that
\begin{eqnarray*}
\Delta_{AC|B}(\rho)=S(\rho||\sigma).
\end{eqnarray*}
Assume state $\rho_1$ is the closest $C$-incoherent state to $\rho$, namely,
\begin{eqnarray*}
 C^r_{AB|C}(\rho)=S(\rho||\rho_1).
\end{eqnarray*}
Suppose $\Pi_C=\{\Pi_C^i\}$ is the local measurement such that $\rho_1=\sum_i\Pi_C^i\rho\Pi_C^i$. One has $\sigma_1=\sum_i\Pi_C^i\sigma\Pi_C^i$. Since $\mathrm{Tr}(\rho\log\rho_1)=\mathrm{Tr}(\rho_1\log\rho_1)$ and $\mathrm{Tr}(\rho\log\sigma_1)=\mathrm{Tr}(\rho_1\log\rho_1)$, we have
\begin{eqnarray*}
S(\rho||\sigma_1)=S(\rho||\rho_1)+S(\rho_1||\sigma_1).
\end{eqnarray*}
As the relative entropy does not increase under quantum operations, $S(\Lambda(\rho)||\Lambda(\sigma))\leq S({\rho}||{\sigma})$, we have
$S(\rho_1||\sigma_1)\leq S(\rho||\sigma)$. Then
\begin{eqnarray*}
S(\rho||\sigma_1)\leq C^r_{AB|C}(\rho)+\Delta_{AC|B}(\rho).
\end{eqnarray*} 
By the definition of deficit, we have $\Delta_{A|BC}(\rho)\leq S(\rho||\sigma_1)$, and
\begin{eqnarray*}
\Delta_{A|BC}(\rho)-\Delta_{AC|B}(\rho)\leq C^r_{AB|C}(\rho).
\end{eqnarray*}

\end{document}